# PROPOSAL FOR AN ENHANCED OPTICAL COOLING SYSTEM TEST IN AN ELECTRON STORAGE RING

E.G.Bessonov, M.V.Gorbunkov, Lebedev Phys. Inst. RAS, Moscow, Russia,
A.A.Mikhailichenko, Cornell University, Ithaca, NY, U.S.A.

**Abstract**

We are proposing to test experimentally the new idea of Enhanced Optical Cooling (EOC) in an electron storage ring. This experiment will confirm new fundamental processes in beam physics and will demonstrate new unique possibilities with this cooling technique. It will open important applications of EOC in nuclear physics, elementary particle physics and in Light Sources (LS) based on high brightness electron and ion beams.

## 1. INTRODUCTION

Emittance and the number of stored particles $-N$ in the beam determine the principal parameter of the beam, its Brightness what can be defined as $B = N/\gamma\varepsilon_x \gamma\varepsilon_z \gamma\varepsilon_s$, where each $\gamma\varepsilon_{x,z,s}$ stands for invariant emittance associated with corresponding coordinate. Beam cooling reduces the beam emittance (its size and the energy spread) in a storage ring and therefore improves its quality for experiments. All high-energy colliders and high-brilliance LS's require intense cooling to reach extreme parameters. Several methods for the particle beam cooling are in hand now: (i) radiation cooling, (ii) electron cooling, (iii) stochastic cooling, (iv) optical stochastic cooling, (v) laser cooling, (vi) ionization cooling, and (vii) radiative (stimulated radiation) cooling [1-3]. Recently a new method of EOC was suggested [4-7] and in this proposal we discuss an experiment which might test this method in an existing electron storage ring having maximal energy ~ 2.5 GeV, and which can also function down to energies of ~100-200 MeV.

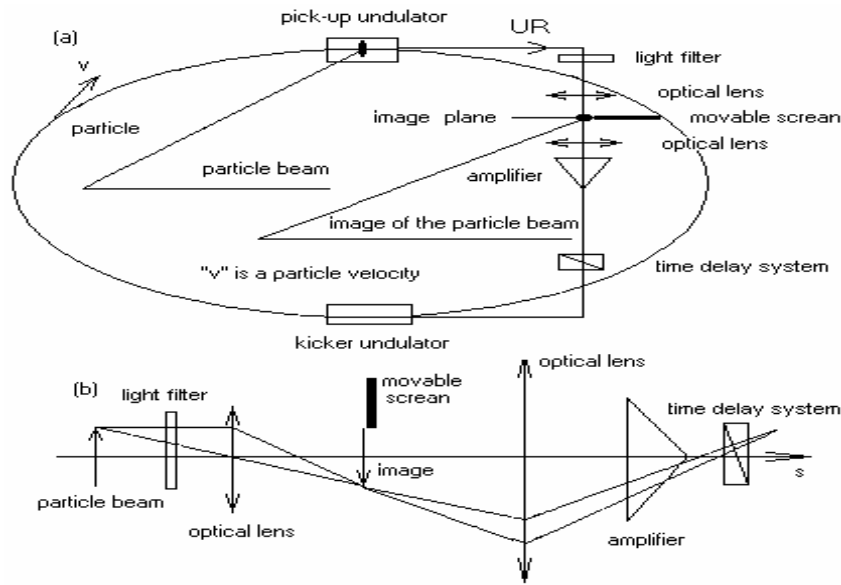

Figure1: The scheme of the EOC of a particle beam (a) and unwrapped optical scheme (b)

EOC [4] appeared as the symbiosis of enhanced emittance exchange and Optical Stochastic Cooling (OSC) [8-10]. These ideas have not yet been demonstrated. At the same time the ordinary Stochastic Cooling (SC) is widely in use in proton and ion colliders. OSC and EOC extend the potential for fast cooling due to bandwidth. EOC can be successfully used in Large Hadron Collider (LHC) as well as in a planned muon collider.

The EOC in the simpiest case of two dimensional cooling in the longitudinal and transverse x-planes is based on one pickup and one or more kicker undulators located at a distance determined by the betatron phase advance $\psi_x^{bet} = 2\pi(k_{p,k} + 1/2)$ for first kicker undulator and $\psi_x^{bet} = 2\pi k_{k,k}$ for the next ones, where $k_{ij} = 0, 1, 2, 3,\ldots$ is the whole numbers. Other elements of the cooling system are the optical amplifier (typically Optical Parametric Amplifier i.e. OPA), optical filters, optical lenses, movable screen(s) and optical line with variable time delay (see Fig.1). An optical delay line can be used together with (or in some cases without) isochronous pass-way between undulators to keep the phases of particles such that the kicker undulator decelerates the particles during the process of cooling [6], [7].

## 2. TO THE FOUNDATIONS OF ENHANCED OPTICAL COOLING

The total amount of energy carried out by undulator radiation (UR) emitted by electrons traversing an undulator, according to classical electrodynamics, is given by

$$E_{tot}^{cl} = \tfrac{2}{3} r_e^2 \overline{B^2} \beta^2 \gamma^2 L_u , \qquad (1)$$

where $r_e = e^2/m_e c^2$ is the classical electron radius; $e$, $m_e$ are the electron charge and mass respectively; $\overline{B^2}$ is an averaged square of magnetic field along the undulator period $\lambda_u$; $\beta = v/c$ is the relative velocity of the electron; $\gamma = E/m_e c^2$ is the relativistic factor; $L_u = M\lambda_u$ is the length of the undulator; and $M$ is the number of undulator periods. For a planar harmonic undulator $\overline{B^2} = B_0^2/2$, where $B_0$ is the peak of the undulator field. For a helical undulator $\overline{B^2} = B_0^2$. The spectral distribution of the first harmonic of UR for $M \gg 1$ is given by [11]

$$dE_1^{cl}/d\xi = E_1^{cl} f(\xi) \qquad (0 \leq \xi \leq 1), \qquad (2)$$

where $E_1^{cl} = E_{tot}^{cl}/(1+K^2)^2$, $f(\xi) = 3\xi(1 - 2\xi + 2\xi^2)$, $\xi = \lambda_{1,\min}/\lambda_1$, $\lambda_{1\min} = \lambda_1|_{\theta=0}$, $\int f(\xi)d\xi = 1$, $K = e\sqrt{\overline{B^2}}\lambda_u/2\pi m_e c^2$ is the deflection parameter, $\lambda_1 = \lambda_u(1 + K^2 + \vartheta^2)/2\gamma^2$ is the wavelength of the first harmonic of the UR, $\vartheta = \gamma\theta$; $\theta$ is the azimuthtal angle between the vector of electron average velocity in the undulator and the undulator axis.

Electrons have effective resonant interaction in the field of the kicker undulator only with that part of their undulator radiation wavelets (URW) emitted in the pickup undulator if the frequency bands and the angles of the electron average velocities are selected in the ranges

$$\left(\frac{\Delta\omega}{\omega}\right)_C = \frac{1}{2M}, \qquad (\Delta\vartheta)_C = \sqrt{\frac{1+K^2}{M}} \qquad (3)$$

nearby maximal frequency and to the axes of both pickup and kicker undulators. Optical filters which are tuned up to the maximal frequency of the first harmonic of the UR can be used for this selection. In this case screens must select the URWs emitted at angles $\vartheta_{URW} < (\Delta\vartheta)_C$ to the pickup undulator axis both in horizontal and vertical directions before they enter optical amplifier (to do away with the unwanted part of URWs loading OPA). In this case the angle between the average electron velocity vector in the undulator and the undulator axis will be small:

$$(\Delta\vartheta)_e < (\Delta\vartheta)_C. \qquad (4)$$

Below we suggest that the optical system of EOC selects a portion of URWs, emitted in this range of angles and frequencies, by filters, diaphragms and/or screens. This condition limits the precision $\delta\psi_{x,z}^{bet}$ of the phase advance $\psi_{x,z}^{bet}$ determined by the equation

$$(\delta\theta)_{x,z}^{bet} < (\Delta\theta)_C , \qquad (5)$$

where $(\delta\theta)_{x,z}^{bet} = (2\pi A_{x,z,bet}/\lambda_{x,z,bet})\sin(\delta\psi_{x,z}^{bet})$ is the change of the angle between the electron average velocity and the axis of the kicker undulator owing to an error in the arrangement of undulators, $A_{x,z,bet}$ is the amplitude of the betatron oscillations of the electron in the storage ring, in the smooth approximation $\delta\psi_{x,z}^{bet} = 2\pi\Delta s/\lambda_{x,z,bet}$, $\Delta s$ is the displacement of the kicker undulator from optimal position, $\lambda_{x,z,bet}$ is the length of the period of betatron oscillations.

The number of the photons in the URW emitted by electrons in suitable cooling frequency and angular ranges (3) is defined by the following formula (see Appendix 1)

$$N_{ph} = \frac{\Delta E_1^{cl}}{\hbar\omega_{1\max}} = \pi\alpha\frac{K^2}{1+K^2}, \qquad (6)$$

where $\Delta E_1^{cl} = (dE_1^{cl}/d\omega)\Delta\omega = 3E_{tot}/2M(1+K^2)^2$, $\omega_{1\max} = 2\pi c/\lambda_{1\min}$, $\alpha = e^2/\hbar c \cong 1/137$ [11]. Filtered URWs must be amplified and directed along the axis of the kicker undulator.

If the density of energy in the URWs has a Gaussian distribution with a waist size $\sigma_w > \sigma_{x,z}^e$, $Z_R > L_u/2$, the R.M.S. electric field strength $E_w^{cl}$ of the wavelet of the length $2M\lambda_{1\min}$ in the kicker undulator defined by the expression

$$E_w^{cl} = \sqrt{\frac{\Delta E_1^{cl}}{M\sigma_w^2\lambda_{1\min}}} = \frac{\sqrt{2}r_e\gamma^2\sqrt{B^2}}{(1+K^2)^{3/2}\sqrt{M}\sigma_w}. \qquad (7)$$

where $\sigma_{x,z}^e$ are the electron beam dimensions, $Z_R = 4\pi\sigma_{w,c}^2/\lambda_{1\min}$ is the Rayleigh length. If $\sigma_{x,z}^e < \sigma_{w,c}$, $\sigma_W = \sigma_{w,c}$, the R.M.S. electric field strength $E_w^{cl}$ of the wavelet becomes

$$E_w^{cl} = \frac{4\sqrt{2\pi}\, r_e\, \gamma^3\sqrt{B^2}}{L_u(1+K^2)^2} \qquad (8)$$

where $\sigma_{w,c} = \sqrt{L_u\lambda_{1\min}/8\pi}$ is the waist size corresponding to the Rayleigh length $Z_R = L_u/2$.

Note that electric field values (7), (8) do not take into account quantum nature of emission of URWs in a pickup undulator. They are valid for $N_{ph} \gg 1$. Such case can be realized only for heavy ions with atomic number $Z > 10$ and for deflection parameter $K > 1$. If, according to classical electrodynamics, $N_{ph} < 1$, then it means that in the reality, according to quantum theory, one photon is emitted with the energy $\hbar\omega_{1,\max}$ and with the probability $p_{em} = N_{ph} < 1$. In this case the electric field strength is determined by the replacement of the energy $\Delta E_1^{cl}$ on $\Delta E_1^q = \Delta E_1^{cl} \cdot N_{ph}^{-1} = \hbar\omega_{1,\max}$ in (7) and the frequency of the emission of photons $f_{ph} = f \cdot p_{em} = f \cdot N_{ph} < f$, where $f$ is the revolution frequency of the electron in the storage ring.

If the number of electrons in the URW sample is $N_{e,s}$, then URW emitted by an electron $i$ in pickup undulator and amplified in OPA decrease the amplitudes of betatron and synchrotron oscillations of this electron in the kicker undulator. Other $N_{e,s} - 1$ electrons emit URWs including $(N_{e,s}-1)N_{ph}$ non-synchronous (for the electron $i$) photons, which are amplified by OPA and together with noice photons of the OPA increase the amplitudes of oscillations of the electron $i$. If the number of non-synchronous photons in the sample $N_{ph,\Sigma} = (N_{e,s}-1)N_{ph} + N_n < 1$, where $N_n$ - is the number of noise photons in the URW sample at the amplifier front end [6], [7], then the jumps of the closed orbit and the electric field strengths are determined by the replacement of the energy $\Delta E_1^{cl}$ on $\Delta E_1^q$ in (7), (8) and the frequency of the emission of photons is $f_{em} = f \cdot p_\Sigma = f \cdot N_{ph,\Sigma} < f$. In the opposite case, $N_{ph,\Sigma} > 1$, the electric field strengths are determined by the replacement of the energy $\Delta E_1^{cl}$ on $\Delta E_1^{q\Sigma} = N_{ph,\Sigma} \cdot \hbar\omega_{1,\max}$ in (7) and the frequency of the emission of photons is $f$.

In our case $N_{e,s} = 2M\lambda_{1,\min}N_{e,\Sigma}/\sigma_{s,0}$, $N_{e\Sigma}$ stands for the number of electrons in the bunch, $\sigma_{s,0}$ is the initial length of the electron bunch.

The maximum rate of energy losses for the electron in the fields of the kicker undulators and amplified URW is

$$P_{loss}^{max} = -eE_w^{cl} L_u \beta_{\perp m} f \Phi(N_{ph}) N_{kick} \sqrt{\alpha_{ampl}} \bigg|_{\sigma_w = \sigma_{w,c}} = \frac{8\pi\sqrt{\pi} e^2 f \Phi(N_{ph}) N_{kick} K^2 \sqrt{\alpha_{ampl}}}{(1+K^2)\lambda_{1,min}}, \quad (9)$$

where $\beta_\perp = K/\gamma$; $N_{kick}$ is the number of kicker undulators (it is supposed that electrons are decelerated in these undulators); and $\alpha_{ampl}$ is the gain in the optical amplifier. The function $\Phi(N_{ph})|_{N_{ph}\ll 1} = \sqrt{p_{em}} = \sqrt{N_{ph}}$, $\Phi(N_{ph})|_{N_{ph}\gg 1} = 1$ takes into account the quantum nature of the emission (the frequency of emission of photons $\sim f \cdot N_{ph}$ and the electric field strength $\sim 1/\sqrt{N_{ph}}$). It follows, that quantum nature of the photon emission in undulators leads to the decrease of the maximum average rate of energy losses for electrons in the fields of the kicker undulator and amplified URW by the factor $\Phi(N_{ph})|_{N_{ph}\ll 1} = \sqrt{N_{ph}} \simeq 9.3$.

The damping times for the longitudinal and transverse degrees of freedom are

$$\tau_{s,EOC} = \frac{6\sigma_{E,0}}{|P_{loss}^{max}|}, \qquad \tau_{x,EOC} = \tau_{s,EOC} \frac{\sigma_{x,0}}{\sigma_{x_\eta,0}} = \frac{6\beta^2 E_s \sigma_{x,0}}{|P_{loss}^{max}| \eta_{x,kick}}, \quad (10)$$

where $\sigma_{E,0}$ is the initial energy spread of the electron beam, $P_{loss}$ stands for the power losses (9), $\sigma_{x,0}$ is the initial radial beam dimension determined by betatron oscillations, $\eta_{x,kick} \neq 0$ is the dispersion function in the kicker undulator, $\sigma_{x_\eta,0} = \eta_{x,kick} \beta^{-2} (\sigma_{E,0}/E)$. Note that the damping time for the longitudinal direction does not depend on $\eta_{x,kick}$ and one for the transverse direction is inverse to $\eta_{x,kick}$. Factor 6 in (10) takes into account that the energy spread for cooling is $2\sigma_{E,0}$, electrons does not interact with their URWs every turn (screening effect) and that the jumps of the electron energy and closed orbit in general case lead to lesser jumps of the amplitude of synchrotron and betatron oscillations [6].

The equilibrium spread in the positions of the closed orbits $\left(\sqrt{\overline{x_\eta^2}}\right)_{eq}^{EOC}$, the spread of betatron amplitudes $\left(\sqrt{\overline{A_x^2}}\right)_{eq}^{EOC}$ and corresponding beam dimensions $\sigma_{x_\eta}^{EOC}$, $\sigma_x^{EOC}$ determined by EOC are

$$\sigma_{x,eq}^{EOC} = \sigma_{x_\eta,eq}^{EOC} = \left(\sqrt{\overline{A_x^2}}\right)_{eq}^{EOC}/\sqrt{2} = \left(\sqrt{\overline{x_\eta^2}}\right)_{eq}^{EOC}/\sqrt{2}\bigg|_{N_{ph,\Sigma}>1} = \frac{1}{2\sqrt{2}}(N_{e,s}-1+N_n/N_{ph})|\delta x_\eta^1|, \quad (11)$$

where $\delta x_\eta^1 = \eta_x \beta^{-2}(\Delta E_{loss}^{max}/E)$ is the jump of the electron closed orbit determined by the energy jump $\Delta E_{loss}^{max} = P_{loss}^{max}/fN_{ph}$ of the electron in the fields of the kicker undulator and its amplified URW (corresponds to one-photon/mode or one-photon/sample at the amplifier front end).

The equilibrium relative energy spread of the electron beam

$$\sigma_{E,eq}^{EOC}/E = \beta^2 \sigma_{x_\eta,kick}^{EOC}/\eta_{x,kick} \quad (12)$$

Note that jumps of closed orbits $\delta x_\eta^1 \sim 1/E$. That is why the electron bunch dimensions (11) at the same number of particles in the sample and relativistic factor are much higher ions one. As a sequence of small electron charge the number of photons in the URWs $N_{ph} = 1.15 \cdot 10^{-2} \ll 1$, 87% of UWRs are empty of synchronous photons and every URW has $N_{ph,\Sigma} > 1$ non-synchronous photons. That is why the contribution of noise photons for electrons is greater ($N_n/N_{ph} \simeq 87 N_n$) then for heavy ions.

The power transferred from the optical amplifier to electron beam is

$$P_{ampl} = \varepsilon_{sample}^{cl} \cdot f \cdot N_{e\Sigma} + P_n, \quad (13)$$

where $\varepsilon_{sample}^{cl} = \hbar\omega_{1,max} N_{ph}\alpha_{ampl}$ is the average energy in a sample, $P_n$ is the noise power. This is the maximal limit for the power corresponding to the case if all electrons are involved in the cooling process simultaneously (screening is absent and the amplification time interval of the amplifier $\Delta t_{ampl}$ is higher then the time duration of the electron bunch $\Delta t_b$).

The initial phases $\varphi_{in}$ of electrons in their URWs radiated in the pickup undulator and transferred to the entrance of the kicker undulator(s) depend on their energies and amplitudes of betatron oscillations. If we assume that synchronous electron enter the kicker undulator together with their URW at the decelerating phase corresponding to the maximum decelerating effect, then the initial phases for other electrons in their URWs will correspond to deceleration as well, if the difference of their closed orbit lengths between undulators remains $\Delta s < \lambda_{1,min}/2$. In this case the amplitudes of betatron oscillations, the transverse horizontal emittance of the beam in the smooth approximation and the energy spread of the beam at zero amplitude of betatron oscillations of electrons must not exceed the values

$$A_x \ll A_{x,\lim} = \sqrt{\lambda_{1\min}\lambda_{x,bet}/\pi}, \qquad \varepsilon_x < 2\lambda_{1\min}, \qquad \frac{\sigma_E}{E} < (\frac{\sigma_E}{E})_{\lim} = \frac{\lambda_{1\min}}{2L_{p,k}}\frac{\beta^2}{\eta_{c,l}}, \qquad (14)$$

where $L_{p,k}$ is the distance between pickup and kicker undulators along the synchronous orbit, $\eta_{c,l} = -d\ln T_{p,k}/d\ln p$ is the local slippage factor between the undulators, $p$ is the momentum of an electron, $T_{p,k} = L_{p,k}/c\beta$ is the pass by time between pickup and kicker undulators. In accordance with the betatron phase advance $\psi_x^{bet} = 2\pi(k_{p,k}+1/2)$; the value $L_{p,k} = \lambda_{x,bet}(k_{p,k}+1/2)$, where $\lambda_{x,bet} = C/v_x$ is the wavelength of betatron oscillations, C is the circumference of the ring, $v_x$ is the betatron tune.

The third equation in (14) can be overcome if the isochronous bend or bypass between undulators will be used. In some cases controllable variable in time optical delay-line can be used to change in *situ* the length of the light pass-way between the undulators during the cooling cycle to keep the decelerating phases of electrons in the kicker undulator in the process of cooling [6], [7].

Below we investigate this case in more details. The difference $\Delta t$ in the propagation time of the URW and the traveling time $T_{p,k}$ of the electron between pickup and kicker undulators depends on initial conditions of electron's trajectory which can be expressed as

$$c\Delta t |_{\gamma\gg 1} = c_t - \int_{s0}^{s}\frac{x}{\rho}d\tau = c_t - x_0\int_{s0}^{s}\frac{C}{\rho}d\tau - x_0'\int_{s0}^{s}\frac{S}{\rho}d\tau - \frac{\Delta E}{\beta^2 E}\int_{s0}^{s}\frac{D}{\rho}d\tau,$$

where $x = x_\beta + x_\eta$, $x_0 = x_{0\beta} + x_{0\eta}$, $x_\beta$ is the deviation of the electron from its closed orbit, $x_\eta$ is the deviation of the closed orbit itself from synchronous one, $x_{0\beta}$ and $x_{0\eta}$ stand for appropriate deviations at location $s = s_0$. Two eigenvectors called sine-like S(z) and cosine-like C(z) trajectories and $\rho$ stands for the local bending radius, $c_t$ is a constant which is determined by the optical delay line. Basically vectors S(z), C(z) describe the trajectories with initial conditions like $x_0'(s_0) = 0$; $x(s,s_0) = x_0 \cdot C(s,s_0)$ and $x_0(s_0) = 0$; $x(s,s_0) = x_0' \cdot S(s,s_0)$, where $s_0$ corresponds to the longitudinal position of the pick up. So the transverse position of the particle has the form [12]

$$x(s) = x_0 \cdot C(s,s_0) + x_0' \cdot S(s,s_0) + D(s,s_0) \cdot (\Delta E/\beta^2 E)$$

where $\Delta E = E - E_d$ is the deviation of the electron energy from the dedicated energy $E_d$ and dispersion D defined as

$$D(s,s_0) = -S(s,s_0) \cdot \int_{s_0}^{s} \frac{C(\tau)}{\rho(\tau)} d\tau - C(s,s_0) \cdot \int_{s_0}^{s} \frac{S(\tau)}{\rho(\tau)} d\tau,$$

Dispersion $D(s,s_0)$ describes transverse position of the test particle having relative momentum deviation from equilibrium as big as $\Delta p/p$, while its initial values of transverse coordinates at $s=s_0$ are zero. So full expression for transverse position of particle comes to form

$$x(s) = x_{0\beta} \cdot C(s,s_0) + x'_{0\beta} \cdot S(s,s_0) + [\eta_{x,0} \cdot C(s,s_0) + \eta'_{x,0} \cdot S(s,s_0) + D(s,s_0)] \frac{\Delta E}{\beta^2 E},$$

where $\eta_x$ describes periodic solution for dispersion in damping ring (slippage factor) and $x_{0\beta}, x'_{0\beta}$ marks pure betatron part in transverse coordinate; $\eta_{0,x}, \eta'_{0,x}$ stand for its value at location of pickup kicker. So the time difference becomes

$$c\Delta t = c_t - R_{51}(s,s_0) \cdot x_0 - R_{52}(s,s_0) \cdot x'_0 - R_{56}(s,s_0) \frac{\Delta E}{\beta^2 E} \cong c_t + cT_{p,k} \eta_{c,l} \frac{\Delta E}{\beta^2 E}, \quad (15)$$

where we neglected terms responsible for the betatron oscillations (i.e. $R_{51}=0$, $R_{52}=0$).

In general case $\eta_{c,l} \neq 0$ the initial phase of an electron in the field of amplified URW propagating through kicker undulator, according to (15), $\varphi_{in} = \omega_{1,max} \Delta t \neq 0$ and the rate of the energy loss

$$P_{loss} = -|P_{loss}^{max}| \sin(\varphi_{in}) \cdot f(\Delta E), \quad (16)$$

where $f(\Delta E) = 1 - |\varphi_{in}(\Delta E)|/2\pi M$, if $|\varphi_{in}| \leq 2\pi M$ and $f(\Delta E) = 0$ if $|\varphi_{in}| > 2\pi M$. The function $f(\Delta E)$ takes into account that electron with some energy $E_d$ and its URW enter kicker undulator simultaneously at the phase $|\varphi_{in}| = 0$ and passing together all undulator length zero rate of the energy loss if $c_t = 0$. Electrons having the energies $E \neq E_d$ so they and their URWs enter the kicker undulator non-simultaneously with different phases, travel together shorter distance in the undulator under smaller rate of the energy change.

According to (16), electrons with different initial phases are accelerated or decelerated and gathered at phases $\varphi_{in}^m = \pi + 2\pi m$ ($-M \leq m \leq M$, $m = 0, \pm 1, .. \pm M$) and at energies

$$E_m = E_d + \frac{(2m+1)\pi \beta^2}{\omega_{1,max} T_{p,k} \eta_{c,l}} E_d = E_d [1 + \frac{(2m+1)\lambda_{1,min} \beta^2}{2 L_{p,k} \eta_{c,l}}], \quad (17)$$

if RF accelerating system is switched off (see Fig.2).

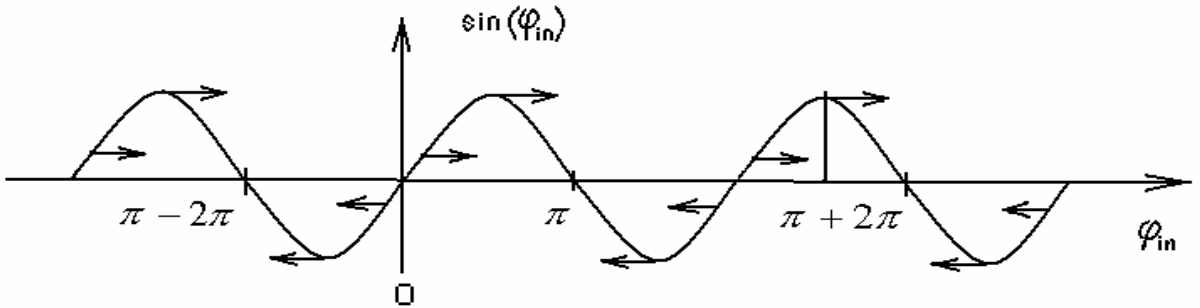

Figure 2: In the EOC scheme electrons are grouping near the phases $\varphi_{in} = \pi + 2\pi m$ (energies $E_m$)

The energy gaps between equilibrium energy positions have magnitudes given by

$$\delta E_{gap} = E_{m+1} - E_m = \frac{\lambda_{1,min}}{L_{p,k} \eta_{c,l}} \beta^2 E_d. \quad (18)$$

Note that the energy gap (18) is 2 times higher the limiting energy spread of the beam at zero amplitude of betatron oscillations of electrons (14).

The power loss $P_{loss}$ is the oscillatory function of energy $|E - E_d|$ with the amplitude linearly decreasing from the maximum value $|P_{loss}^{max}|$ at the energy $E = E_d$ to a zero one at the energy $|E - E_d| \geq M \cdot \delta E_{gap}$. If the RF accelerating system is switched off, the electron energy falls into the energy range $|E - E_d| < M \cdot \delta E_{gap}$, excitation of synchrotron oscillations by non-synchronous photons can be neglected then the electron energy is drifting to the nearest energy value $E_m$. The variation of the particle's energy looks like it produces aperiodic motion in one of $2M$ potential wells located one by one. The depth of the wells is decreased with their number $|m|$. If the delay time in the optical line is changed, the energies $E_m$ and the energies of particles in the well are changed as well. In this case particles stay in their wells if their maximal power loss satisfies the condition

$$|P_{loss}^{max}| > |dE_m / dt + P_{loss}^{ext}|, \qquad (19)$$

where $|P_{loss}^{ext}|$ stands for the external power losses determined by synchrotron radiation.

## 3. VARIANTS OF OPTICAL COOLING

Depending on the local slippage factor and coefficients $R_{51}$, $R_{52}$ and $R_{56}$ in (14), different variants of optical cooling can be suggested.

1. The local slippage factor $\eta_{c,l} = 0$, betatron oscillations are absent, dispersion function in the pickup undulator $\eta_{x,pickup} \neq 0$. In this case $\delta t = const$ and the initial phase for all electrons can be installed $\varphi_{in} = \pi/2$. It corresponds to electrons arriving kicker undulator in decelerating phases of theirs URWs under maximum rate of energy loss. In this case electrons will be gathered near to the synchronous electron if a moving screen opens the way only to URWs emitted by electrons with the energy higher than synchronous one. This is the case of an EOC in the longitudinal plane based on isochronous bend and screening technique.

If electrons develop small betatron oscillations (betatron oscillations introduce phase shift less than $\pi/2$, (see (14)), then the electron beam will be cooled in transverse and longitudinal directions simultaneously. If the dispersion function value in the pickup undulator $\eta_x = 0$ or if the synchrotron oscillations of electrons are small (no selection in longitudinal plane) then the cooling in the transverse direction only takes place ($\eta_{x,kick} \neq 0$).

2. The scheme of OSC can be used at $\eta_{c,l} = 0$ [8]. In this scheme the pickup undulator is a quadrupole one and kicker undulator is ordinary one. They have the same period. The magnetic field in the quadrupole undulator is increased with the radial coordinate $x$ by the low $B(x) \cong G \cdot x$ and changes the sign at $x = 0$, where $G$ stands for the gradient. The phase of the emitted URWs changes its value on $\pi$ at $x = 0$ as well. That is why electrons are grouped around synchronous orbit in the ring where they do not emit URWs. The deflection parameter in the quadrupole undulator increased with the radial coordinate ($K \propto |B(x)|$) and so the emitted wavelength also $\lambda_{1\min} \cong \lambda_u \cdot (1 + K^2(x))/2\gamma^2$. As the resonance interaction of URW and the electron emitted the URW is possible in the kicker undulator only if deflection parameters of undulators are near the same, this opens a possibility for initial selection of amplitudes in the pickup undulator. So the cooling can be arranged for some specific amplitude of synchrotron oscillations only. The continuous resonance interaction and cooling is possible if the magnetic field of kicker undulator is decreased in time for cooling process. Electrons having other than resonance synchrotron amplitude do not interact with cooling system. Betatron oscillations in this scheme must introduce the phase shift less then $\pi/2$ as well. This can be arranged by proper zeroing cos and sin-like trajectory integrals [13].

The scheme with two quadrupole undulators (the pickup one and the kicker one) described in [13]. In this case the second quadrupole undulator decreases the amplitudes of synchrotron oscillations for positive deviations $x_{\eta,pickup} > 0$ (we choose the conditions when electrons are decelerated in their URWs if $x_{\eta,pickup} > 0$ and betatron amplitudes are neglected ($x_{\beta,pickup} = x_{\beta,kick} = 0$)) and increases them for negative ones $x_{\eta,pickup} < 0$ as it experience deceleration again (the phases of URWs change their value on $\pi$ at $x = 0$ and simultaneously the electron will pass the kicker undulator at opposite magnetic field). In this case to cool electron beam additional selection of URWs can be used by the screen (cut off URWs emitted by electrons at negative deviations $x$). Another scheme which can be used, based on truncated undulator with the magnetic field of the form $B(x)|_{x>0} \simeq G \cdot x$ and $B(x)|_{x<0} \simeq 0$. Such undulator can be linearly polarized one with upper or down array of magnetic poles. It was used in the undulator radiation experiments in circular accelerators [14].

3. If $\eta_{c,l} \neq 0$, betatron oscillations of electrons introduce phase shift $<< \pi/2$ and the energy gaps have the magnitude $\delta E_{gap} = (3 \div 5) \cdot \sigma_{E,0}$, then transit-time method of OSC based on two identical undulators can be used [9]. In this case the main part of electrons including tail electrons will be gathered at the energy $E_s$, if the energy $E_0 = E_m|_{m=0}$ was chosen equal to $E_s$. Decreasing of the beam dimensions leads to decrease of the rate of cooling. In this case a time depended local slippage factor $\eta_{c,l}(t)$ can be used to decrease the energy gap for cooling process and to increase the rate of cooling.

4. If $\eta_{c,l} \neq 0$, the energy gaps between equilibrium energy positions have the magnitudes $\delta E_{gap} \cong (3 \div 5) \cdot \sigma_{E,0} / M$, RF accelerating system is switched off then electrons are gathered at phases $\varphi_{in}^m$ and energies $E_m$ independently on amplitudes of betatron oscillations. If the screen overlap URWs emitted by electrons at negative deviation from one having minimum energy $E_{min}$ and optical system change the delay time of the rest URWs to move the energies $E > E_{min} + (3 \div 5) \cdot \sigma_{E,0}$ to the energy $E_{min}$ then electrons loose their energy and amplitudes of betatron oscillations until their energy takes minimum one. Cooling takes place according to the scheme of the EOC.

5. For this variant $\eta_{c,l} \neq 0$, the energy gaps between equilibrium energy positions have the magnitudes $\delta E_{gap} << \sigma_{E,0} / M$, the RF accelerating system of the storage ring is switched off, the screen absorbs the URWs emitted by electrons at a negative deviation of theirs position from the synchronous one in the radial direction, energy layers are located at positive deviations from synchronous one outside the energy spread of the beam and optical system change the delay time of the URWs to move the energy layers to the synchronous energy. Then the energy layers capture small part of electrons of the beam first and electrons with smaller energy are captured increasingly and loose their energy and betatron amplitudes until reaching the minimum energy $E_{min}$ allowed in the beam. So the cooling process takes place. This process can be repeated. In this case the energy jump $\delta E_{loss}^{max} = |P_{loss}^{max}|/f N_{ph}$ of the electron in the kicker undulator must be less than the energy gap $\delta E_{gap}$ determined by synchronous photons (18) and non-synchronous photons in the URWs having higher energy jumps $\delta E_{non-syn}^{max} = \delta E_{loss}^{max} \sqrt{N_{ph,\Sigma}}$ at $N_{ph,\Sigma} > 1$. That is why the next condition must be fulfilled

$$\delta E_{loss}^{\max} \sqrt{N_{ph,\Sigma}} \mid_{N_{ph,\Sigma} > 1} < \delta E_{gap} = \frac{\lambda_{l,\min}}{L_{p,k} \, \eta_{c,l}} \beta^2 E_d \,. \tag{20}$$

Otherwise electrons can jump over the energy layers and cooling will not be effective.

If the RF accelerating system of the storage ring is switched on, the average energy loss per turn $\Delta E_{loss}^{turn} = |P_{loss}^{\max}|/f$ is higher than the energy loss $|eV_0(\sin\phi - \sin\phi_s)|$ of the electron in the RF accelerating system of the storage ring or

$$|\sin\phi - \sin\phi_s| < \delta E_{loss}^{turn}/eV_0, \tag{21}$$

then the energy of electrons is drifting to the nearest energy value $E_m$ and EOC takes place. Here $V_0$ is the amplitude of the RF accelerating voltage, $\phi_s$ is the synchronous phase determined by the equation $\sin\phi_s = \Delta E_{SR,s}/eV_0 = V_s/V_0$, $\Delta E_{SR,s} = P_{SR,s}/f = eV_s$ is the energy loss per turn, $P_{SR,s} = \frac{2}{3}cr_e^2 \overline{B^2} \gamma^2 \simeq 2.77 \cdot 10^3 \gamma^4 / R_s \overline{R_s}$ eV/sec is the average power of the synchrotron radiation emitted by the electron in the ring. The value $P_{SR,s}/f = 5.8 \cdot 10^{-7} \gamma^4 / R_s$ eV/turn.

To keep the condition (21) satisfied, the range of RF phases of particles $2|\phi - \phi_s|$ interacting with their URWs must be limited by the value $2|\phi_{c,1} - \phi_s|$ determined by equality in (21). This can be done by using OPA with short amplification time interval $\Delta t_{ampl,1}$ corresponding to the range of phases $\omega_{RF} \cdot \Delta t_{ampl} < 2|\phi_{c,1} - \phi_s|$ and by overlapping the center of this time interval with synchronous particle, where $\omega_{RF} = 2\pi f_{RF}$, $f_{RF}$ is the frequency of the RF accelerating system of the ring. The last condition is equivalent to

$$l_{ampl}^{laser} < l_{ampl,1} = 2c |\phi_{c,1} - \phi_s| / \omega_{RF}, \tag{22}$$

where $l_{ampl,1}$ is the length of the amplified laser bunch. In this case $2M+1$ electron ellipses appear around synchronous phase in the longitudinal plane. The amplitudes of synchrotron oscillations are determined by the energies $E_m$ which are moved to synchronous energy if the optical system changes the delay time of the URWs. The condition (20) must be fulfilled if the RF accelerating system of the storage ring is switched on as well.

The electrons will be gathered effectively on elliptic trajectories having maximum energies $E_m$ (17) if conditions (20), (21) are fulfilled, and the deviation $\delta E = E - E_m$ of the electron energies from $E_m$ in the process of interaction are small. The last condition can be overcame by limiting the amplification time by interval $\Delta t_{ampl,2}$ and the corresponding length $l_{ampl,2} = c\Delta t_{ampl,2}$ to

$$l_{ampl}^{laser} < l_{ampl,2}^{laser} = \frac{\sigma_{s,0}}{2\sqrt{2}} \sqrt{\frac{\delta E_{gap}}{\sigma_{E,0}}}, \tag{23}$$

where $\sigma_{s,0}$ is the initial length of the electron bunch. Above we suggested that electrons are moving along elliptical trajectories $\Delta E = \pm \sigma_E \sqrt{1 - s^2/\sigma_s^2}$ and interact with URWs at the top energies $\sim E_m$ in the region of the energy deviations $\delta E = E - E_m = \delta E_{gap}/8$.

Multiple processes of excitation of synchrotron oscillations by non-synchronous and noise photons will increase the widths of the electron ellipses and to transfer electrons from one ellipse to another. They can be neglected if the equilibrium energy spread (12) of the beam is less then the energy gap (18) or

$$\sigma_{E,eq}^{EOC} < \delta E_{gap} \tag{24}$$

Damping time (10) in variant 5 will be increased $2\sigma_{s,0}/l_{ampl}$ times:

$$\tau_s^{EOC} = \frac{12\sigma_{E,0}}{|P_{loss}^{max}|} \frac{\sigma_{s,0}}{l_{ampl}} \qquad \tau_x^{EOC} = \frac{12\beta^2 E_s \sigma_{x,0}}{|P_{loss}^{max}| |\eta_{x,\,kick}|} \frac{\sigma_{s,0}}{l_{ampl}}. \qquad (25)$$

The variant 5 permits to avoid any changes in the existing lattice of the ring (isochronous bend, bypass). It works easier for existing ion storage rings (see Appendix 2).

The screen permits us to select in pickup undulator electrons with positive deviations of both betatron and synchrotron oscillations, and such a way to produce effective cooling both in the transverse and longitudinal direction (we suggested $\eta_x \neq 0$ in pickup and kicker undulators in this case). Using the number of kicker undulators $N_{kick} > 1$ permits to cool the beam either in two directions or in the transverse or in the longitudinal directions only by selecting corresponding distances between kicker undulators [4], [6].

## 4. OPTICAL SYSTEM FOR THE EOC SCHEME

UR of an electron gets its well known properties only after the electron passed the undulator and UR is considered in far zone. The lens located near the pickup undulator can strongly influence to the UR properties if its focus is inside the undulator [15].

For effective cooling of an electron beam in a storage ring, parameters of the beam under cooling and the optics of EOC system must fulfill certain requirements.

1. The URW, emitted in a pickup undulator must be filtered and passed through the laser amplifier.

2. In variant 1 each electron in a beam should enter kicker undulator simultaneously with its amplified URW emitted in a pickup undulator and to move in decelerating phase of this URW. For the test electron of a beam (for example, for the synchronous electron with zero amplitude of betatron oscillations) this requirement is satisfied by equating the propagation time of the URW with the traveling time of the electron between undulators. Conditions (14) are necessary for other electrons of the beam to get decelerating phases of their URWs in this case.

3. Each electron in the beam should enter the kicker undulator with its URW emitted in the pickup undulator near the center of this wavelet in transverse direction. This requirement will be satisfied if the transverse sizes of all URWs in the kicker undulator are overlapped.

The rms transverse size of one URW at the distance $l$ from the pickup undulator is equal to $\sigma_w \simeq (\Delta\theta)_c (l + M\lambda_u/2)$ (assuming that radiation is emitted from the center of the undulator). At the distance $l$ from the undulator the R.M.S. transverse size of the beam of emitted URWs is equal $\sigma_{w,b} \simeq d + (\Delta\theta)_c M\lambda_u/2 + (\Delta\theta)_c \cdot l$, where $d$ is the transverse size of the electron beam. If the optical screen opens the way to radiation from the part of the electron beam only, so the only small angles to the undulator axis $\Delta\vartheta < (\Delta\vartheta)_C$ passed through, then at the distance from the end of the undulator $l = l_c$, where

$$l_c = \frac{d}{(\Delta\theta)_c} + \frac{M\lambda_u}{2} \qquad (26)$$

URWs will be overlapped, and the transverse size of the beam of the selected wave packets will be equal $2d + (\Delta\theta)_c M\lambda_u$ (see. Fig. 3). If the beam of URWs passed optical lenses, movable screen, the optical amplifier, optical delay and injected into the kicker undulator then electrons under cooling will hit theirs URWs in the transverse plane if the transverce dimensions of the electron beam in the undulator are less then $2\sigma_{w,b}$.

4. Generally the angular resolution of an electron bunch by an optical system is

$$\delta\varphi_{res} \cong 1.22 \frac{\lambda_{1\min}}{D}, \qquad (27)$$

where $D$ is diameter of the first lens. This formula is valid, if elements of the source emit radiation which is distributed uniformly in a large solid angle. In our case only a fraction of the lens affected by radiation, as $D > \sigma_{w,b}$. That is why the effective diameter $D_{eff} = \sigma_{w,b}$ must be used in (27). At the distances $l > l_c$ the size $\sigma_{w,b} \simeq (\Delta\theta)_c l$, so the space resolution of the optical system is $\delta x_{res} \cong \delta \varphi_{res} l$ or

$$\delta x_{res} \cong 1.22 \lambda_{1\min} / (\Delta\theta)_c = 0.86 \sqrt{\lambda_{1\min} L_u}, \tag{28}$$

where $(\Delta\theta)_c = (\Delta\vartheta)_c / \gamma = (1/\gamma)\sqrt{(1+K^2)/M}$ is the observation angle.

Note that at closer distances $l < l_c$, the spatial resolution is better. More complicated optical system can be used for increase the spatial resolution in this case.

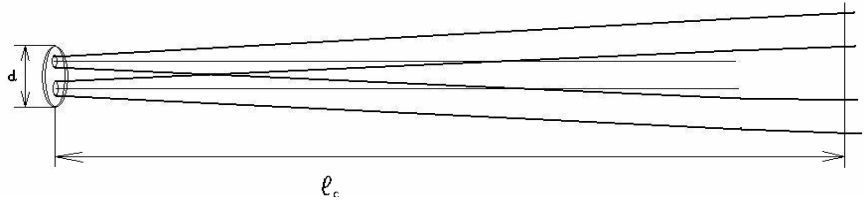

Figure 3: Scheme of URW's propagation.

5. URWs must be focused on the crystal of OPA. For the URW beam, the dedicated optical system with focusing lenses can be used to make the Rayleigh length equal to the length of the crystal (typically ~1 cm) for small diameters of the focused URW beam in the crystal (typically ~0.1 mm).

6. The electron bunch spacing in storage rings is much bigger than the bunch length. The same time structure of the OPA must take advantage on this circumstance.

## 5. USEFUL EXPRESSIONS FROM THE THEORY OF CICLIC ACCELERATORS

The equilibrium value of relative energy spread of the electron beam in the isomagnetic lattice of the storage ring determined by synchrotron radiation (SR) described by expression

$$\frac{\sigma_E^{eq,SR}}{E} = \sqrt{\frac{55 \Lambda_c \gamma^2}{32\sqrt{3} R_s \Im_s}} \simeq 6.2 \cdot 10^{-6} \frac{\gamma}{\sqrt{R_s \Im_s}}, \tag{29}$$

where $\Lambda_c = \hbar/mc \simeq 3.86 \cdot 10^{-11}$ cm, $R_s$ is the equilibrium bending radius of the synchronous electron, in the smooth approximation $\Im_s = 4 - \alpha_c \overline{R_s}/R_s$ is a coefficient determined by the magnetic lattice ($0 < \Im_s < 3$), $\alpha_c$ is the momentum compaction factor, $\overline{R_s}$ is the equilibrium averaged radius of the storage ring [16].

For small synchrotron oscillations the equilibrium length of the electron bunch is

$$\sigma_s^{eq,SR} = \frac{\alpha_c c}{\Omega} \frac{\sigma_E^{eq,SR}}{E}, \tag{30}$$

where $\Omega = \omega_{rev}\sqrt{h\alpha_c e V_0 \cos\phi_s / 2\pi E_s}$ is the synchrotron frequency, $\omega_{rev} = 2\pi f$, $h$ is an accelerating RF voltage harmonic order.

The equilibrium radial synchrotron and betatron beam dimensions are

$$\sigma_{x_\eta}^{eq,SR} = \eta_x \frac{\sigma_E^{eq,SR}}{E}, \qquad \sigma_x^{eq,SR} = \gamma\sqrt{\frac{55\sqrt{3}}{96} \frac{\Lambda_c \overline{R_s}}{\Im_x} F} = 6.2 \cdot 10^{-6} \gamma \sqrt{\frac{\overline{R_s} F}{\Im_x}}, \tag{31}$$

where in the smooth approximation $\Im_s = \alpha_c \overline{R_s}/R_s - 1$ is a coefficient determined by the magnetic lattice ($\Im_x + \Im_s = 3$), $F \sim 1$ is a coefficient [16], [17].

The damping time in the storage ring determined be synchrotron radiation in the bending magnets and comes to the following values ($\Im_z = 1$)

$$\tau_{x,z,s}^{SR} = \frac{2E}{P_{SR,s}\mathfrak{I}_{x,z,s}} = \frac{3R_s\overline{R_s}}{cr_e\gamma^3\mathfrak{I}_{x,z,s}} = 355\frac{R_s\overline{R_s}}{\gamma^3\mathfrak{I}_{x,z,s}} \text{ [sec]}. \tag{32}$$

The maximum deviation of the energy from its synchronous one is

$$\frac{(\Delta E)_{sep}}{E} = \pm \beta \sqrt{\frac{eV_0}{\pi h \eta_c E}[(\pi - 2\phi_s)\sin\phi_s - 2\cos\phi_s]}, \tag{33}$$

where $\eta_c = \alpha_c - 1/\gamma^2$ is the storage ring slippage factor of the ring.

For the ordinary storage ring lattice (without local isochronous bend or bypass between undulators) the natural local slippage factor

$$\eta_{c,l} \simeq L_{p,k}\eta_c/C. \tag{34}$$

## 6. EXAMPLE

Below we consider an example of one dimensional EOC of an electron beam in the transverse x-plane in strong focusing storage ring like Siberia-2 (Kurchatov Institute Atomic Energy, Moscow) having maximal energy 2.5 GeV [18]. The magnetic system of the ring is designed with so-called separate functions. The lattice consists of six mirror-symmetrical superperiods, each containing an achromatic bend and two 3 m long straight sections. For the functionality, the half of the superperiod arranged with two sections. The first one, comprising the quadrupoles F,D,F and two bending magnets is responsible for the achromatic bend and high $\beta_x$, $\beta_y$ functions in the undulator straight section. The second part, comprising quadrupoles D,F,D and dispersion free straight section, allows to change the betatron tunes, without disturbing the achromatic bend. Main parameters of the ring, the electron beam in the ring, pickup and kicker undulators and Optical Parametric Amplifier are presented in Tables 1 - 4.

After single bunch injection in the storage ring the energy 100 MeV is establishd for the experiment and the beam is cooled by synchrotron radiation damping (see section 5). In this case the energy spread and the beam size acquire equilibrium values in ~40 seconds (see Table1). The equilibrium energy spread is equal to $\sigma_E^{eq,SR}/E = 3.94\cdot 10^{-5}$, the length of the bunch $\sigma_s^{eq,SR} = 2.32$ cm at the amplitude of the accelerating voltage $V_0 = 73$ V, the synchronous voltage $V_s = 1.89$ V, the radial emittance $\epsilon_x^{eq,SR} = 1.25\cdot 10^{-6}E^2[GeV] = 1.25\cdot 10^{-8}$ cm rad, the radial betatron beam dimension at pickup undulator $\sigma_x^{eq,SR} = 4.61\cdot 10^{-2}$ mm.

Following synchrotron radiation damping the amplitudes of radial betatron oscillations $\sigma_{x,0}$ are artificially excited to be suitable for resolution of the electron beam in the experiment with EOC (see Table 2). The amplitudes of synchrotron oscillations must stay damped to work with short electron bunches and short duration of amplification OPAs.

In the variants of the example considered below the optical system resolution of electron beam, according to (28), is $\delta x_{res} = 1.9$ mm at $\lambda_{1,\min} = 2\cdot 10^{-4}$ cm, $M\lambda_u = 240$ cm. It yields that effective EOC in this case is possible if the beam under cooling has total size in the pickup undulator $\sigma_{x,tot} > 2.0$ mm. We accepted the initial energy spread $\sigma_{E,0} = \sigma_E^{eq,SR} = 3.94\cdot 10^{-5}E$, the dispersion beam size $\sigma_{x_\eta,0} = 3.15\cdot 10^{-2}$ mm, the length of the electron bunch $\sigma_{s,0} = 2\sigma_s^{eq,SR} = 4.64$ cm, its transverse size at pickup undulator $\sigma_{x,0} = 4$ mm, the laser amplification length $l_{ampl}^{laser} = 1.5$ mm (duration 0.5 ps), the radial betatron beam size in kicker undulator $\sigma_{x,0} = 1$ mm, the URW beam size $\sigma_{w,b} = 2$ mm. We took the number of electrons at the orbit $N_{e,\Sigma} = 5\cdot 10^4$. In this case the number of electrons in the URW sample is $N_{e,s} = 129.5$, the number of non-synchronous photons in the sample is $N_{ph,\Sigma} = 2.5$ for the case of one noice photon at the OPA front end. In this storage ring the natural local slippage factor (34) is $\eta_{c,l} = \eta_c L_{p,k}/C \simeq \alpha_c L_{p,k}/C = 4.45 \cdot 10^{-3}$, the energy gap (18) is $\delta E_{gap} = 0.62$ keV.

We consider EOC in the transverse plane. In this case the dispersion beam size $\sigma_{x_\eta,0} \ll \delta x_{res}$ and that is why there is no selection of electrons in the longitudinal plane. That is why in order to prevent heating in the longitudinal plane by energy jumps determined by both synchronous and non-synchronous photons in the URWs, two kicker undulators are used which produce zero total energy jump [4], [6]. Note that the purpose of this experiment is to check physics of optical cooling. At the same time cooling in the transverce plane is important for heavy ions in RHIC, LHC.

We accept the distance between pickup and first kicker undulator along the synchronous orbit $L_{p,k} = 72.27$ m ($\psi_x^{bet} = 9\pi$, $k_{p,k} = 4$). It corresponds to the installation of undulators in the first and seventh straight sections which are located at a distance 72.38 m (we count off pickup undulator). Second kicker undulator is located on the same distance from the first one. Optical line is tuned such a way that electrons are decelerated in the first kicker undulator and accelerated in the second one.

The URWs have the number of the photons emitted in the pickup undulator (see Table 3) $N_{ph} = 1.15 \cdot 10^{-2}$ per electron in the frequency and angular ranges (3) suitable for cooling. The limiting amplitude of betatron oscillations (14) is $A_{x,\lim} = 3.2$ mm. The energy spread of the beam limited by the separatrix is $\Delta E_{sep}/E = 3.3 \cdot 10^{-4}$. The electric field strength at the first harmonic of the amplified URW in the kicker undulator is $E_w^{cl} \cong 2.06 \cdot 10^{-3}$ V/cm. The power loss for the electron passing through one kicker undulator together with its amplified URW comes to $P_{loss}^{\max} = 2.03 \cdot 10^6$ eV/sec if the amplification gain of OPA is $\alpha_{ampl} = 10^7$ (see Tables 2, 4). This power loss corresponds to the maximal energy jumps $\Delta E_{loss}^{\max} = 73$ eV and the average energy loss per turn $\Delta E_{loss}^{turn} = 0.84$ eV/turn. The jump of the closed orbits is $\delta x_\eta^1 = 5.8 \cdot 10^{-5}$ cm. Below we will consider two variants of EOC.

*1. The variant 1 (section 3).* For the parameters presented above the cooling time for the transverse coordinate, according to (10), comes to $\tau_{x,EOC} = 18.5$ msec. SR damping time ~ 40 sec is much bigger (see Table 1). The average power transferred from the optical amplifier to electron beam (13) is $\overline{P}_{ampl} = 0.061$ mW. It is determined by the power of the URWs (0.036 mW) and noise average power (41) equal to 0.025 mW. We adopted one-photon/mode (one-photon/sample) at the amplifier front end corresponding to pulse noise power at the amplifier front end $P_{n,0} = \hbar\omega c/M\lambda = 4 \cdot 10^{-7}$ W, $P_n = 2$ W at the gain $G_0 = 10^7$, used $\hbar\omega \simeq 0.5$ eV. We took into account that the amplification time interval of the amplifier $\Delta t_{ampl}$ is less than the revolution period by a factor of $C/c\Delta t_b = C/2\sigma_{s,0} = 8.27 \cdot 10^4$ times.

Necessary conditions for selection of electrons must be created: high beta function in the pickup undulator (to increase the transverce dimensions of the bunch for selection of electrons with positive deviations from closed orbit), the isochronous bend between undulators. We believe that the lattice of the ring is flexible enough to be changed in nesessary limits by analogy with those presented in [19].

The number of electrons in the bunch is enough to detect them in the experiment and to neglect intrabeam scattering.

Note that if one kicker undulator is used in the scheme of two-dimensional EOC and the beam resolution is high $\delta x_{res} < 10^{-2}$ mm, the equilibrium relative energy spread, the spread of closed orbits, the longitudinal, dispersion and radial betatron beam dimensions determined by EOC, according to (11), are equal to $\sigma_E^{eq,EOC}/E_s = 5.56 \cdot 10^{-5}$, $\sigma_{s,0}^{eq,EOC} = 2.63$ cm, $\sigma_{x_\eta}^{eq,EOC} = \sigma_x^{eq,EOC} =$ $4.46 \cdot 10^{-2}$ mm. It follows that if the number of electrons in the bunch $N_{e,\Sigma} < 5 \cdot 10^4$ then their influence on the equilibrium dimensions of the bunch can be neglected, longitusinal dimensions and the energy spread stay small and the radial betatron beam dimensions determined by EOC are high degree decreased. In reality the equilibrium synchrotron and betatron bunch dimensions will be much higher. This is the consequence of the finite beam resolution in pickup undulator. That is why we use two kicker undulators to keep longitudinal bunch dimensions small to exclude the excitation of longitudinal oscillations by multiple energy jumps. The situation could be better if we had effective OPA at the wavelength $\lambda_{1,\min} \simeq 3 \cdot 10^{-3}$ cm, i.e. about one order less. Shorter undulator can be used as well.

*2. The variant 5 (section 3).* The variant 5 requires easier tuning of the lattice for the arrangement of the local small slippage factor between undulators. In the case of one-dimensional EOC, using two kicker undulators, the multiple processes of excitation are not essential because of the excitation of the synchrotron oscillations in this case is absent or unessential and that is why there is no need in the small local slippage factor. In this case the initial phase $\varphi_{in}(E, A_x)$ of the electron in the field of amplified URW propagating through the kicker undulator, according to (15) is the function of both the energy (which is a constant in this variant of EOC) and the amplitude of betatron oscillations. The amplitudes of betatron oscillations will increase or decrease depending on their initial phases until they reach the equilibrium amplitudes determined, in the smooth approximation, by the expression $A_{x,m} = \sqrt{2m \cdot \lambda_{1\min} \lambda_{x,bet} / \pi}$ (generelised expression (14)) corresponding to the phases $\varphi_m = 2m\pi$. Variable in time optical delay-line can be used to change in *situ* the length of the light passway between the undulators during the cooling cycle to move the initial phases to $\varphi_{in} \simeq 2m\pi + \pi/2$ for production of the optimal rate of decrease the amplitudes of betatron oscillations of electrons in the fields of amplified URW and the kicker undulator. The damping time for radial betatron oscillations, according to (25), is $\tau_{x,EOC} = 0.57$ sec.

Note that in the case of two-dimensional EOC using one kicker undulator, according to (18), the energy gaps between equilibrium energy positions have the magnitudes $\delta E_{gap} = 0.62$ keV. They are higher than the energy jumps of electrons in kicker undulator $\Delta E_{loss}^{\max} = 73$ eV and the energy jumps $\Delta E_{loss}^{\max} \sqrt{N_{ph,\Sigma}} = 115$ eV determined by the non-synchronous photons in the URWs (see condition (20)). The conditions (22), (23) or $l_{ampl}^{laser} < l_{ampl,1,2}$ limit the length of the laser URWs by the values $l_{ampl,1} = 1.69$ cm, $l_{ampl,2} = 0.32$ cm. The accepted laser amplification length $l_{ampl}^{laser} = 1.5$ mm is enough to satisfy these conditions. The damping time for radial betatron oscillations, according to (25), is $\tau_{x,EOC} = 1.14$ sec. This damping time is less then one for synchrotron radiation damping (see Table1). The equilibrium energy spread determined by EOC is about 6-10 times higher then the energy gap. It follows that the local slippage factor between undulators, according to (24), must be decreased by a factor higher then 10. Unfortunately the resolution of the electron beam will not permit to reach the equilibrium energy spread and cooling in the transverse plane in this case will be less then heating in the longitudinal one.

## 7. CONCLUSIONS

We have shown in this paper, that test of EOC is possible in the 2.5 GeV electron strong focusing storage ring tuned down to the energy ~ 100 MeV. Electron beam can be cooled in transverse direction. The damping time is much less than one determined by synchrotron radiation. So the EOC can be identified by the change of the damping rate of the electron beam. Variant of cooling is found, which permits to avoid any changes in the existing lattice of the ring (for production of isochronous bend, bypass). It can work for existing ion storage rings as well (see Appendix 2). Three short undulators in this variant installed in the storage ring have rather long periods and weak fields. They can be manufactured at low cost.

The cooling of a relatively small number of electrons (one bunch, $N_{e,\Sigma} = 5 \cdot 10^4$) is considered in this proposal in attempt to avoid strong influence of the non-synchronous photons on the equilibrium energy spread of the beam. The intrabeam scattering effects could be overcame as well. Optical amplifier suitable for the EOC - so called Optical Parametric Amplifier - suggested as a baseline of experiment, must have moderate gain and power. We have chosen the wavelength of the OPA equal 2 mkm as the OPA technique is more developed for these wavelengths. At the present times the OPAs having amplification gain ~$10^8$ and the power >1 W are fully satisfy requirements for this experiment (See Appendix 3). Usage of OPAs with shorter wavelength will permit to increase the spatial resolution by the optical system and the degree of cooling of the beam.

We have predicted that the maximum rate of energy loss for electrons in the fields of the kicker undulator and amplified URW calculated in the framework of classical electrodynamics is $\sqrt{N_{ph}} \simeq 9.3$ times lesser then one taking into account quantum nature of the photon emission in undulators. Quantum aspects of the beam physics will be checked in the proposed test experiment. It is suggested that the scheme based on optical line with variable delay time will be tested as well.

Authors thank A.V.Vinogradov and Yu.Ja.Maslova for useful discussions.
Supported by RFBR under grant No 05-02-17162a, 05-02-17448a, by the Program
of Fundamental Research of RAS, subprogram "Laser systems" and by NSF.

**Table 1. Parameters of the ring:**

| | |
|---|---|
| The maximal energy of the storage ring | $E_{max}$= 2.5 GeV |
| The energy for the experiment | $E_{exp}$=100 MeV |
| Relativistic factor for the experiment | $\gamma \cong 200$ |
| Circumference | C=124.13 m |
| Bending radius | $R = 490.54$ cm |
| Average radius | $\bar{R} = 1976$ cm |
| Frequency of revolution | $f = 2.42 \cdot 10^6$ Hz |
| Harmonic number | $h = 75$ |
| RF frequency | $f_{RF} = 181.14$ MHz |
| Energy loss determined by SR | $P_{\gamma,SR} / f = 1.89$ eV/turn |
| The amplitude of the accelerating voltage at $E_{exp}$ | $V_0 = 73$ V |
| The synchronous phase | $\varphi_s \simeq 0.026$ |
| Radial tune | $v_x = 7.731$ |
| Vertical tune | $v_z = 7.745$ |
| Momentum compaction factor | $\alpha_c = 7.6 \cdot 10^{-3}$ |
| Dispersion function at the pickup and the kicker undulator | $\eta_x = 80$ cm, |
| Radial beta function in pickup undulator | $\beta_x = 17$ m |
| Radial beta function in kicker undulator | $\beta_x = 1.7$ m |
| Vertical beta function | $\beta_z = 6$ m |
| Patrician coefficients $\Im_z, \Im_x, \Im_s$ | 1, 0.97, 2.03 |
| Damping times by SR at $E_{exp}$ $\tau_z, \tau_x, \tau_s$ | 43.1; 44.4; 21.23 sec |
| The length of the period of betatron oscillations | $\lambda_{x,bet} = 16.06$ m |

| Slippage factor of the ring | $\eta_c = \alpha_c$ |
| Local slippage factor of the ring | $\eta_{c,l} = 0.58 \cdot \alpha_c$ |
| Frequency of synchrotron oscillations at $E_{exp} = 100$ MeV | $\Omega = 1.6 \cdot 10^{-3} f$. |

**Table 2. Initial parameters of the electron beam in the ring:**

| | |
|---|---|
| Number of electrons at the orbit | $N_{e,\Sigma} = 5 \cdot 10^4$, |
| Number of electron bunches being cooled | 1 |
| Relative energy spread | $\sigma_{E,0}/E = 3.94 \cdot 10^{-5}$, |
| Betatron beam size at pickup undulator ($\beta_x = 32$ m) | $\sigma_{x,0} = 4$ mm, |
| Betatron beam size at kicker undulator ($\beta_x = 2$ m) | $\sigma_{x,0} = 1$ mm, |
| Dispersion beam size | $\sigma_{x_\eta,0} = 3.15 \cdot 10^{-2}$ mm, |
| Total beam size at pickup undulator | $\sigma_{x,tot} = \sqrt{(\sigma_{x_\eta,0})^2 + (\sigma_{x,0})^2} = 4$ mm, |
| The length of the electron bunch | $\sigma_{s,0} = 2.32$ cm, |

**Table 3. Parameters of pickup and kicker undulators:**

| | |
|---|---|
| Magnetic field strength | $\sqrt{B^2} = 1338$ Gs, |
| Undulator period | $\lambda_u = 8$ cm, |
| Number of periods | $M = 30$, |
| Deflection parameter | $K=1$. |

**Table 4. Optical Parametric Amplifier**

| | |
|---|---|
| Number of Optical Parametric Amplifiers | 2 |
| Total gain | $\alpha_{ampl} = 10^7$ |
| The wavelength of amplified URWs | $\lambda_{1,min} = 2 \cdot 10^{-4}$ cm |
| The characrerisitic URW waist size | $\sigma_{w.c} = 0.77$ mm |
| The URW beam waist size | $\sigma_w = 2$ mm |
| The duration of the amplification time of the OPA | 5 psec ($l_{ampl} = 1.5$ mm) |
| The frequency of the amplified cycles | $f_{ampl} = f = 2.42 \cdot 10^6$ Hz |

**Appendix 1**

Spectral-angular distribution of the UR energy emitted by the relativistic electron in the pickup undulator on the harmonic $n$ is

$$\frac{\partial^2 E_n}{\partial \omega \, \partial o} = \frac{M}{\omega_1} \frac{\partial E_n}{\partial o} \sin c^2 \sigma_n(\omega, \vartheta), \quad (35)$$

where $\partial E_n / \partial o$ is the angular distribution of the energy of the UR emitted in the unit solid angle $do$ at the angle $\theta$ to it's axis, $\sin c\, \sigma_n = \sin \sigma_n / \sigma_n$, $\sigma_n = \pi n M(\omega - \omega_n)/\omega_n$, $\omega_n = n\omega_1$ is the frequency of the $n$-th harmonic of UR. [20] - [22]. For the helical undulator

$$\frac{\partial E_n}{\partial o} = \frac{e^2 M n^2 \omega_1^3(\vartheta) \beta_\perp^2 F_n(K,\vartheta)}{c\Omega^2} = \frac{6 E_{tot} \gamma^2}{\pi} \frac{n^2 F_n(K,\vartheta)}{(1+K^2+\vartheta^2)^3},$$

$F_n(K,\vartheta) = J_n^{'2}(n\chi) + (\frac{1+K^2-\vartheta^2}{2K\vartheta})^2 J_n^2(n\chi)$, $E_{tot} = 4\pi e^2 \Omega M \gamma^2 K^2/3c$, $\Omega = 2\pi c/\lambda_u$, $J_n, J_n^{'}$ is the Bessel function and it's derivative, $\chi = 2K\vartheta/(1+K^2+\vartheta^2) < 1$, $\beta_\perp = K/\gamma$. The number of photons emitted in the undulator on the harmonic *n* in the solid angle $do = \pi d\vartheta^2/\gamma^2$ and frequency band $d\omega$ is determined by the relation

$$N_{ph,n} = \frac{\alpha n^2 M^2 K^2 \lambda_u}{\pi c} \iint \frac{F_n(K,\vartheta)}{(1+K^2+\vartheta^2)^2} \operatorname{sinc}^2 \sigma_n(\omega,\vartheta) d\omega do. \qquad (36)$$

If the considered frequency band $\Delta\omega/\omega \ll 1/M \ll 1$ then we can neglect the angular dependence of the first multiplier and the frequency dependence of the value $\operatorname{sinc} \sigma_n$ in (36). In this case the value $\operatorname{sinc} \sigma_n$ determine the range of angles of the UR (3). The increasing of the frequency band will lead to the increase of the angular range. Taking the frequency band $\Delta\omega$ out of the integral and taking into account that $d\vartheta^2 = (2\Omega\gamma^2/\pi M\omega_1)d\sigma_n$, $\operatorname{sinc}\sigma_n = \pi\delta(\sigma_n)$ we can transform (36) to (6) for $\Delta\omega/\omega = 1/2M$  $\vartheta = 0$ and $n=1$.

**Appendix 2**
Below we investigate the possibility of lead ions cooling (Z=82) in the storage ring LHC based on using the version 5 (section 3) of EOC. We take the example 2 considered in [7]. In this case the energy $E = 1.85\cdot 10^{14}$ eV, ($\gamma = 953$) the slippage factor of the ring $\eta_c \simeq \alpha_c = 3.23\cdot 10^{-4}$, the amplitude of the RF accelerating voltage $V_0 = 16$ MV, the RMS bunch length 8 cm, RF frequency $f_{RF} = 400$ MHz, synchrotron frequency $\Omega = 23$ Hz, circumference C=2665888.3 cm, harmonic order h=35640, RMS relative energy spread $\sigma_{E,0}/E = 1.1\cdot 10^{-4}$, $\sigma_E^{in} = 2.04\cdot 10^{10}$ eV, $\lambda_{x,bet} = 415$ m, RF bucket half-height $\Delta E_{sep}/E = 4.43\cdot 10^{-4}$, $\Delta E_{sep} = 8.19\cdot 10^{10}$ eV. We take the distance between pickup and kicker undulators $L_{p,k} = 1453$ m (k=3), synchrotron radiation energy loss per ion per turn $P_{SR,s}/f = 1.2\cdot 10^4$ eV.

For the parameters of the undulators [7] the energy loss per turn is $\Delta E_{loss}^{\max} = 3\cdot 10^5$ eV, $\sigma_{E,0}/M = 1.7\cdot 10^9$ eV (M=12), the gap between equilibrium energy positions is $\delta E_{gap} = 1.97\cdot 10^8$ eV, $\lambda_{1,\min} = 5\cdot 10^{-5}$ cm. It follows that $\Delta E_{loss}^{\max} \ll \delta E_{gap}$, that is the condition (20) is satisfied. In the case of ions the equation (21) must include $eZ\cdot V_0$ instead of $eV_0$. In this example $\delta E_{loss}^{\max}/eZV_0 = 2.2\cdot 10^{-4}$. It follows the laser amplification length $l_{ampl,2} = 2.78$ mm. By this choice the condition (21) is satisfied as well. To cool the ion beam in the transverse plane and to keep the magnetic lattice unchanged one pickup and two kicker undulators must be used.

**Appendix 3**
The principle of OPG is quite simple: in a suitable nonlinear crystal, a high frequency and high intensity beam (the *pump* beam, at frequency $\omega_p$) amplifies a lower frequency, lower intensity beam (the *signal* beam, at frequency $\omega_s$); in addition a third beam (the *idler* beam, at frequency $\omega_i$, with $\omega_i < \omega_s < \omega_p$) is generated (In the OPG process, signal and idler beams play an interchangeable role, we assume that the signal is at higher frequency, i.e., $\omega_s > \omega_i$)..
In the interaction, energy conservation

$$\hbar\omega_p = \hbar\omega_s + \hbar\omega_i$$

is satisfied; for the interaction to be efficient, also the momentum conservation (or phase matching) condition

$$\hbar\mathbf{k}_p = \hbar\mathbf{k}_s + \hbar\mathbf{k}_i$$

where $\mathbf{k}_p$, $\mathbf{k}_s$, and $\mathbf{k}_i$ are the wave vectors of pump, signal, and idler, respectively, must be fulfilled. The signal frequency to be amplified can vary in principle from $\omega_p/2$ (the so-called degeneracy condition) to $\omega_p$, and correspondingly the idler varies from $\omega_p/2$ to 0; at degeneracy, signal and idler have the same frequency. In summary, the OPG process transfers energy from a high-power, fixed frequency pump beam to a low-power, variable frequency signal beam, thereby generating also a third idler beam.

The signal and idler group velocities $v_s$ and $v_i$ (GVM – group velocity mismatch) determine the phase matching bandwidth for the parametric amplification process. Let us assume that perfect phase matching is achieved for a given signal frequency $\omega_s$ (and for the corresponding idler frequency $\omega_i = \omega_p - \omega_s$. If the signal frequency increases to $\omega_s + \Delta\omega$, by energy conservation the idler frequency decreases to $\omega_i - \Delta\omega$. The wave vector mismatch can then be approximated to the first order as

$$\Delta k \cong -\frac{\partial k_s}{\partial \omega_s}\Delta\omega + \frac{\partial k_i}{\partial \omega_i}\Delta\omega = \left(\frac{1}{v_{gi}} - \frac{1}{v_{gs}}\right)\Delta\omega.$$

The gain bandwidth of an OPA can be estimated using the analytical solution of the coupled wave equations in the slowly varying envelope approximation and assuming flat top spatial and temporal profiles and no pump depletion. The intensity gain ($G$) and phase ($\varphi$) of the amplified signal beam are given in [23] by

$$G = 1 + (\gamma L)^2 \left(\frac{\sinh B}{B}\right)^2, \quad \varphi = \tan^{-1}\frac{B\sin A\cosh B - A\cos A\sinh B}{B\cos A\cosh B - A\sin A\sinh B}, \quad (37)$$

where $A = \Delta kL/2$, $B = \sqrt{(\gamma L)^2 - (\Delta kL/2)^2}$, and $\gamma = $ gain coefficient $= 4\pi d_{eff}\sqrt{I_p/2\varepsilon_0 n_p n_s n_i c \lambda_s \lambda_i}$, $\Delta kL = $ phase mismatch $= (\mathbf{k}_p - \mathbf{k}_s - \mathbf{k}_i)L$, where $L$ is the length of amplifier, $d_{eff}$ is the effective nonlinear coefficient, $I_p$ is the pumping intensity.

The full width at half maximum (FWHM) phase matching bandwidth can then be calculated within the large-gain approximation as

$$\Delta\nu \cong \frac{2(\ln 2)^{1/2}}{\pi}\left(\frac{\gamma}{L}\right)^{1/2}\frac{1}{\left|\frac{1}{v_{gs}} - \frac{1}{v_{gi}}\right|}.$$

(38)
Large GVM between signal and idler waves dramatically decreases the phase matching bandwidth; large gain bandwidth can be expected when the OPA approaches degeneracy ($\omega_s \rightarrow \omega_i$) in type I phase matching or in the case of group velocity matching between signal and idler ($v_{gs} = v_{gi}$). Obviously, in this case Eq. (5) loses validity and the phase mismatch $\Delta k$ must be expanded to the second order, giving

$$\Delta\nu \cong 2\frac{(\ln 2)^{1/4}}{\pi}\left(\frac{\gamma}{L}\right)^{1/4}\frac{1}{\left|\frac{\partial^2 k_s}{\partial \omega_s^2} + \frac{\partial^2 k_i}{\partial \omega_i^2}\right|}.$$

For the case of perfect phase matching ($\Delta k = 0$, $B = \gamma L$) and in the large gain approximation ($\gamma L \gg 1$), Eq. (37)(4) simplify to

$$I_s(L) \cong \tfrac{1}{4} I_{s0} \exp(2\gamma L), \qquad I_i(L) \cong \frac{\omega_i}{4\omega_s} I_{s0} \exp(2\gamma L). \qquad (39)$$

Note that the ratio of signal and idler intensities is such that an equal number of signal and idler photons are generated. Equations (39) allow defining a parametric gain as

$$G \cong \frac{1}{4} \exp(2\gamma L)$$

growing exponentially with the crystal length $L$ and with the nonlinear coefficient $\gamma$.

$$G \cong \frac{1}{4} \exp\left(8L\pi d_{eff} \sqrt{I_p/2\varepsilon_0 n_p n_s n_i c \lambda_s \lambda_i}\right) \cong \frac{1}{4} \exp\left((8L\pi d_{eff}/n_s \lambda_s)\sqrt{I_p/2\varepsilon_0 n_p c}\right) \qquad (40)$$

for $n_i \approx n_s$, and $\lambda_i \approx \lambda_s$, $\dfrac{1}{\varepsilon_0 c} = 377 \text{ Ohm}$.

The noise (amplified self emission) power of the optical amplifier is determined by the expression

$$P_n = P_{n,0} G_0, \qquad (41)$$

where $P_{n,0}$ is the noise power at the amplifier front end, $G_0$ is the gain of the amplifier. If the noise power corresponds to one-photon/mode at the amplifier front end then $P_{n,0} = \hbar \omega_{1\max}/\tau_{coh}$ [27], [28], where in our case $\tau_{coh} = 2MT$ is the coherence length, $T = \lambda_{1\min}/c$.

**Example: MgO Periodically Poled Lithium Niobate**

In recent years the development of periodically poled nonlinear materials has enhanced the flexibility and performance of OPAs. In the case of well-studied Periodically Poled Lithium Niobat (PPLN), one can get access to the material's highest effective nonlinearity as well as retain generous flexibility in phase-matching parameters and nonlinear interaction lengths.

Operation near the degeneracy wavelength of 2.128 μm reduces thermal-lens effect because the signal and the idler wavelengths fall within the highest transparency range of Lithium Niobat. For wideband optical parametric amplification, we will use MgO: PPLN with a poling period of 31.1 (31.2) μm, which has high damage threshold and high nonlinear coefficient 16pm/V [25] (17pm/V [26]). To avoid photorefractive damage, thick (~1-2 mm) PPLN crystal was suggested to be kept at elevated temperatures $150^0$C [24]. For the signal wavelength $\lambda_s = 2$ μm, $n_s \approx n_p = 2.1$, and the gain, according to formula (40) comes to

$$G \cong \frac{1}{4} \exp\left(3\sqrt{\frac{I_p}{GW/cm^2} \frac{l}{mm}}\right)$$

For $G=10^7$, $I_p$=1GW/cm$^2$, $l = 5.8$ mm. We propose two-stage crystal amplifier ($l_1 = l_2 = 3.5$ mm). OPA amplifies linearly polarized radiation. That is why the circular polarization of the URWs (if helical undulator is used) in our case must be transformed into linear polarized one before it will be injected in the kicker undulator. Usual quarter wave plate can be used for this purpose in simplest case. Planar undulators can be used as well. Reflective optics can be used for dispersion-free undulator light propagation.